\documentclass[11pt,amssymb,superscriptaddress]{revtex4-1}

\usepackage{graphicx,color}     
\usepackage{subfigure}          
\usepackage{epsfig}

\usepackage{amsthm, amsfonts}
\usepackage{amssymb,letterspace}
\usepackage{amscd, amssymb,latexsym}

\usepackage{bm}         
\usepackage{bbold}

\usepackage{slashed}   

\usepackage{dcolumn}    

\begin{document}

\title{Properties of Holographic Mesons on Dense Medium}

\author{Bum-Hoon Lee}
 \email{bhl@sogang.ac.kr}
 \affiliation{Department of Physics, Sogang University, Seoul, Korea 121-742}
 \affiliation{CQUeST, Sogang University, Seoul, Korea 121-742}

\author{Chanyong Park}
 \email{cyong21@sogang.ac.kr}
 \affiliation{CQUeST, Sogang University, Seoul, Korea 121-742}

\author{Siyoung Nam}
 \email{stringphy@gmail.com}
 \affiliation{Department of Physics, Sogang University, Seoul, Korea 121-742}

\begin{abstract}
\textrm{We study the energy dispersions of holographic light mesons and their decay constants on dense nuclear medium. As the spatial momenta of mesons along the boundary direction increase, both observables of the mesons not only increase but also split according to the isospin charges. The decay constant of the negative meson is more large than that of the positive meson of the same type due to the chemical potentials of the background nucleons.}
\end{abstract}

\maketitle

\tableofcontents

\section{Introduction}

\rmfamily

During the last decade, one of the important developments in theories of the high energy physics is the AdS/CFT correspondence \cite{Maldacena:1997re, Witten:1998qj, Gubser:1998bc}, which provides a map between string theories on Anti-de Sitter(AdS) and conformal field theories(CFT) living on the boundary of the AdS. Motivated by the success of the AdS/CFT correspondence, many generalizations to maps between gravity theories and quantum field theories(QFT), especially quantum chromodynamics(QCD), has been proposed and tested under the notion of the gauge/gravity duality. Usually, QFT calculations are performed by perturbation theory and this makes it inaccessible to capture the physics in the strongly-coupled regime of QFT's. In that sense, the gauge/gravity duality is extremely useful because we use a weakly-coupled gravity theory to describe dynamics of a strongly-coupled QFT. For example, the confinement in a hadronic state is a non-perturbative phenomenon, which makes it difficult to study the physics of it. However, one of the recent applications of the gauge/gravity duality, the so-called AdS/QCD sheds light on studying mesons in the confinement phase.

In this work, we will be interested in the hard-wall AdS/QCD model \cite{Erlich:2005qh, DaRold:2005vr, Erlich:2006hq}. This is a bottom-up approach to QCD whose goal is to reproduce known properties of QCD by choosing an appropriate five-dimensional asymptotic AdS gravity.  In this model, it is possible to incorporate chiral symmetry breaking by a bi-fundamental bulk scalar field containing quark mass and chiral condensate \cite{Klebanov:1999tb}. Moreover, a sharp cut-off, called hard-wall, located at the finite bulk radial coordinate $z=z_{{\rm IR}}$ generates the mass gap and confinement \cite{Polchinski:2001tt, Polchinski:2002jw}. Some static hadronic observables such as masses, decay constants, charge radius, and form factors have been studied and reproduced in the hard-wall AdS/QCD model \cite{Erlich:2005qh, DaRold:2005vr, Brodsky:2006uqa, Grigoryan:2007vg, Grigoryan:2007wn, Hong:2004sa, Kwee:2007dd, Abidin:2008ku, Abidin:2008hn}.

These works can be further generalized to the bottom-up models on dense medium \cite{Lee:2013oya, Lee:2009bya, Park:2009nb, Jo:2009xr, Kim:2011gw, Park:2011zp, Kim:2007xi, Nishihara:2014nva, Nishihara:2014nsa}. The density in the dual boundary field theory implies to introduce a $\mathop{U}(1)$ gauge field in the bulk AdS gravity action. Then, the well-known solution is given by Reissner-Nordstrom AdS (RN AdS) black hole, from which thermal charged AdS(tcAdS) space can be constructed to study confining phase of the dense medium at zero temperature, which is nothing but RN AdS black hole of vanishing mass with a hard-wall \cite{Lee:2009bya}. The inserted hard-wall plays a role of setting up an infrared cutoff of the dual boundary theory in the confining phase and, at the same time, prevents any harm to bulk quantities from the naked singularity. The generalization to the two-flavour $\mathop{U}(2)$ case is also possible where combinations of two Cartans are mapped to the chemical potentials or number densities of u and d quark \cite{Park:2011zp}. Since the geometry corresponds to confining phase of the boundary theory, the fundamental excitations are not quarks but nucleons. Appropriate combinations of such bulk gauge fields were reinterpreted as the density of nucleons and this geometry was used to study meson masses and binding energy of a heavy quarkonium in nuclear matter \cite{Lee:2013oya}.

In this paper, we will work in the same background used in \cite{Lee:2013oya} in order to study the dependence of  dispersion and decay constant of mesons on the background nucleon density and their spatial momenta along the boundary. Like in the mass spectrum, if the spatial momenta increase, the non-trivial splitting and increasing of meson energies are developed due to the flavour charges of the background nucleons. As for decay constant of the mesons, the similar behaviors are observed but the decay constant of negative meson is more large than that of positive meson of the same type.

The paper is organized in the following way. We briefly introduce the tcAdS with a hard-wall and present the Lagrangian of the fluctuations in Sec. II. The numerical analysis of vector mesons is given in Sec. III. The cases of the axial-vector and pions are discussed in Sec. IV and V, respectively. In the final Sec. VI, we summarize the paper.


\section{Backgrounds for Confinement at Zero Temperature}


In the hardwall model of AdS/QCD, either a pure AdS space or its Euclidean version, thermal AdS(tAdS) space, is an useful geometry for studying the physics of confinement phase. Since we want to consider mesons on a dense medium in confining phase, some gauge fields must be included in bulk gravity action. Then, the thermal charged AdS(tcAdS) space can be one of the useful candidates.

We choose bulk gauge groups to be $\mathop{U}(2)_L\times\mathop{U}(2)_R$ to realize the full chiral symmetry of the $N_f=2$ lightest flavors in the dual boundary theory. If we denote the corresponding left and right gauge fields as $L_M=L_M^aT^a$ and $R_M=R_M^aT^a$, the gauge field strengths are defined by
\begin{eqnarray}
F^{(L)}_{MN} &=& \partial_ML_N-\partial_NL_M-i[L_M,L_N]\,,\nonumber\\
F^{(R)}_{MN} &=& \partial_MR_N-\partial_NR_M-i[R_M,R_N]\,,
\end{eqnarray}
where $T^a$ ($a=0,\ldots,3$) is a generator of $\textrm{U(2)}$. tcAdS can be obtained by solving field equations derived from the action
\begin{equation}
S_{BG}=\int d^5x\sqrt{-G}\left[\frac{1}{2\kappa^2}({\cal{R}}-2\Lambda) -\frac{1}{4g^2}\textrm{Tr}\left(F^{(L)}_{MN}F^{(L)MN}+F^{(R)}_{MN}F^{(R)MN}\right)\right]\,,
\end{equation}
where the cosmological constant is $\Lambda=-6/R^2$. For simplicity, we set the radius of the curvature $R$ unity. The gauge coupling $g$ and the gravitational coupling $\kappa$ are related to the rank $N_c$ of the color group via,

\begin{equation}
g=\sqrt{\frac{4\pi^2}{N_c}} ~~~~~~\textrm{and}~~~~~~\kappa=\sqrt{\frac{4\pi^2}{N^2_c}}\,.
\end{equation}

By solving the Einstein equations and the Maxwell equations simultaneously, a general solution is given as a well-known Reissner-Nordstr\"{o}m AdS (RN AdS) black hole. There exist two zero-temperature limits of the RN AdS black hole such as an extremal RN AdS black hole and a tcAdS geometry. To work in confinement phase of dual boundary theory, we introduce a hardwall in the bulk. Then it turns out that, within a hardwall formalism, tcAdS is more preferable to confinement in low density and low temperature regime than the usual extremal RN AdS black hole \cite{Lee:2009bya}. The resulting tcAdS solution merely takes the form of the zero-temperature limit of the RN AdS black hole of vanishing mass with the inserted hardwall, which hides a naked singularity. This hardwall also breaks the scaling symmetry in the bulk, which, in turn, implies the broken conformal symmetry, {\it i.e.} setting up an IR cutoff in the dual boundary theory according to the gauge/gravity duality. It is a well-known fact that four-dimensional QCD in low energy scale is in the phase of confinement. Since the number densities of u and d quarks corresponding to bulk gauge fields are conserved and we are in the confinement phase, then appropriate recombinations of bulk gauge fields may be reinterpreted as those of nucleons. This can be justified by observing that turning off any dense effect reproduces the known mass spectrum and value of decay constant of mesons in \cite{Erlich:2005qh}. So it is convenient to write the background solution in terms of the nucleons, not of the quarks, because we are mainly interested in the confining phase on a dense medium. If we denote $Q_P$ and $Q_N$ as number densities of the proton and the neutron, the total number density and the density difference between the nucleons are defined as, respectively
\begin{equation}
Q~=~Q_P\,+\,Q_N ~~~~{\rm and}~~~ D~=~Q_P\,-\,Q_N\,.
\end{equation}
It is also useful to define a relation $D =\alpha Q$ where the range for $\alpha$ is given by $-1\leq \alpha\leq 1$. If $\alpha =0$, then the background density originates from equal numbers of the protons and the neutrons. A nuclear matter which is composed of only one species, either the proton or the neutron, implies $\alpha=\pm1$. $Q$ and $D$ are essentially connected to the bulk gauge fields by setting $V^0_t\equiv L^0_t=R^0_t$ and $V^3_t\equiv L^3_t=R^3_t$. With these, the non-vanishing gauge fields are given by
\begin{eqnarray}
V^0_t &=& \frac{Q}{\sqrt{2}}\left(2z^2_{IR}-3z^2\right)\,,\nonumber\\
V^3_t &=& \frac{D}{3\sqrt{2}}\left(2z^2_{IR}-3z^2\right)\,,
\end{eqnarray}
and the metric of the geometry is given by
\begin{equation}
ds^2 = \frac{R^2}{z^2}\left(-fdt^2 +\frac{dz^2}{f}+d\vec{x}^2\right)\,,
\end{equation}
where
\begin{equation}
f = 1+\frac{3Q^2\kappa^2}{g^2}z^6 +\frac{D^2\kappa^2}{3g^2}z^6\,.
\end{equation}

We pose to mention two interesting possible backgrounds. The first case corresponds to isospin matter with two flavours \cite{Son:2000xc, Parnachev:2007bc, Kim:2007gq, Aharony:2007uu, Albrecht:2010eg, Lee:2013oya}. When referring to \cite{Lee:2013oya}, this is described by setting $Q_\alpha=0$ in original bulk gauge field written in terms of quarks ($\alpha=u,d$), and results in constant gauge fields, $V^3_t=\sqrt{2}\pi^2(\mu_{{\rm P}}-\mu_{{\rm N}})$ and $V^0_t=0$, in the confining phase where $\mu_{{\rm P}}$ and $\mu_{{\rm N}}$ are isospin chemical potentials of proton P and neutron N, respectively. The mass formula for mesons on the isospin matter shows the clear patterns of splitting dependent on the isospin charges. The second example obtained by setting $Q\neq0$ and $\alpha=0$ (or $D=0$) describes a degenerate case, because the parameter $D$ explains the effect of isospins and $Q$ the one of background densities. This can be easily seen by examining equations of motion with respect to mesons, which will be presented in the latter part of the paper. All the equations for each meson take the same form, solely the one of neutral mesons.

The existence of the hard wall makes the radial coordinate $z$ runs from the boundary $z=0$ to the hard wall $z=z_{IR}$. On this background, we turn on various fluctuations to investigate dispersion relations and decay constants of light mesons in the confining phase. Our starting action is given by
\begin{equation}
S =\int d^5x\sqrt{-G}Tr\left[\left|D_M\Phi\right|^2 +m^2\left|\Phi\right|^2 +\frac{1}{4g^2}\left( F^{(L)}_{MN}F^{(L)MN}+F^{(R)}_{MN}F^{(R)MN}\right)\right]\,,
\end{equation}
where the covariant derivative for a scalar field $\Phi$ is defined by
\begin{equation}
D_M\Phi = \partial_M\Phi-iL_M\Phi+i\Phi R_M\,.
\end{equation}
The fluctuations are defined by
\begin{eqnarray}
L^a_M &=& \bar{V}^a_M +l^a_M\,,\nonumber\\\
R^a_M &=& \bar{V}^a_M +r^a_M\,,\nonumber\\
\Phi &=& \phi\mathbb{1}\,e^{i\sqrt{2}\pi}\,,
\end{eqnarray}
where the bar notation is used for denoting the background gauge fields $V_M^a$ and the lower cases represent the fluctuations of the corresponding fields. The chiral symmetry breaking is described by the scalar field

\begin{equation}
\phi = m_qz\,{{}_2F_1}\left(\frac{1}{6},\frac{1}{2},\frac{2}{3},-\frac{(D^2+9Q^2)}{3N_C}z^6\right) +\sigma z^3\, {}_2F_1\left(\frac{1}{2},\frac{5}{6},\frac{4}{3},-\frac{(D^2+9Q^2)}{3N_C}z^6\right)\,,
\end{equation}
where $m_q$ represents a quark mass and $\sigma$ a chiral condensate. Since we are interested in the $\textrm{SU(2)}$ flavor group and work in the axial-like gauge, then $l^i_\mu$, $r^i_\mu$, and $\pi^i$ can only survive during the expansion of the Lagrangian by fluctuations over the fixed dense background.

Further, we redefine the fields as new ones interpreted as vector mesons $v^i_\mu$ and axial-vector mesons $a^i_\mu$, whose definitions are
\begin{equation}
v^i_\mu=\frac{l^i_\mu+r^i_\mu}{\sqrt{2}} ~~~\textrm{and}~~~a^i_\mu=\frac{l^i_\mu-r^i_\mu}{\sqrt{2}}\,.
\end{equation}

At this stage of the Lagrangian up to quadratic-order fluctuations, there exist coupled terms between the axial-vector and the pion fields, which can be removed by decomposing the axial-vector field into the transverse piece $\bar{a}^i_\mu$ and the longitudinal piece such as
\begin{equation} \label{Decomposition}
a^i_\mu =\bar{a}^i_\mu +\partial_\mu\chi^i\,,
\end{equation}
subject to the divergenceless condition $\partial_\mu\bar{a}^{i\,\mu}=0$. After imposing another ansatz $v^i_t=0=\bar{a}^i_t$, we obtain the final Lagrangian of the quadratic fluctuations
\begin{equation}
{\cal{L}}^{(2)}_2 ~=~ {\cal{L}}_{v}+{\cal{L}}_a+{\cal{L}}_{\pi\chi}\,,
\end{equation}
where the Lagrangians for the vector, the axial-vector, and the pseudo-scalar fluctuations are given by, respectively,

\begin{eqnarray}
4g^2{\cal{L}}_v &=& \partial_Mv^i_n\partial^Mv^{in}+\bar{V}^3_t\bar{V}^{3t}\left(v^i_mv^{im}-v^3_mv^{3m}\right) -2\epsilon_{3ij}\bar{V}^{3t}\partial_tv_m^iv^{jm}\,,\\
4g^2{\cal{L}}_a &=& \partial_M\bar{a}^i_n\partial^M\bar{a}^{in} +\bar{V}^3_t\bar{V}^{3t}\left(\bar{a}^i_m\bar{a}^{im}-\bar{a}^3_m\bar{a}^{3m}\right) +4g^2\phi^2\bar{a}^i_m\bar{a}^{im} -2\epsilon_{3ij}\bar{V}^{3t}\partial_t\bar{a}_m^i\bar{a}^{jm}\,,\nonumber\\
4g^2{\cal{L}}_{\pi\chi} &=& \partial_z\partial_\mu\chi^i\partial^z\partial^i\chi^i +\bar{V}^3_t\bar{V}^{3t}\left(\partial_m\chi^i\partial^m\chi^i-\partial_m\chi^3\partial^m\chi^3\right) +4g^2\phi^2\left[ \partial_\mu\chi^i\partial^\mu\chi^i\right. \nonumber\\
&&\left. +\partial_M\pi^i\partial^M\pi^i +\bar{V}^3_t\bar{V}^{3t}\left(\pi^i\pi^i-\pi^3\pi^3\right)  -2\partial_\mu\chi^i\partial^\mu\pi^i -2\epsilon_{3ij}\bar{V}^{3t}\pi^i\partial_t(\chi-\pi)^j\right]\,. \nonumber
\end{eqnarray}


\section{Dispersion Relation and Decay Constant of the $\rho$ Meson}


We study low energy modes of the vector fluctuations, which are identified as the states of $\rho$ mesons. For the purpose of it, we define
\begin{equation}
\rho^0_m ~=~ v^3_m~~~~~\textrm{and}~~~~~\rho^{\pm}_m ~=~ \frac{1}{\sqrt{2}}(v^1_m\pm i \,v^2_m)\,,
\end{equation}
where the superscript $0$ and $\pm$ represent their isospin charges. In terms of these, the Lagrangian of the vector fluctuation is rewritten as
\begin{equation}
2g^2\cdot{\cal{L}}_v =\frac{1}{2}\,\partial_M\rho^0_n\partial^M\rho^{0n} +\partial_M\rho^+_n\partial^M\rho^{-n}+\bar{V}^3_t\bar{V}^{3t}\rho^+_m\rho^{-m} +i\bar{V}^3_t\left(\rho^+_m\partial^t\rho^{-m}-\rho^-_m\partial^t\rho^{+m}\right)\,.
\end{equation}

It is convenient to work in the momentum representation and we write the four-dimensional Fourier-transforms of  $\rho$-mesons as
\begin{eqnarray}
\rho^0_n(x,z) &=& \int\frac{d^4p}{(2\pi)^4}\,\tilde{\rho}^0_n(p,z)\,e^{i(\vec{p}\,\cdot\,\vec{x}-w_0t)}\,,\nonumber\\
\rho^\pm_n(x,z) &=& \int\frac{d^4p}{(2\pi)^4}\,\tilde{\rho}^\pm_n(p,z)\,e^{i(\vec{p}\,\cdot\,\vec{x}-w_\pm t)}\,,
\end{eqnarray}
where $p\equiv\sqrt{\vec{p}\,^2}$ is the magnitude of the spatial momenta in the boundary space. Then, the equations of motion for the $\rho$ vector mesons are given by
\begin{eqnarray}
0 =&& \partial_z\left(\frac{f}{z}\partial_z\tilde{\rho}^0_n\right) +\frac{1}{zf}\left(w^2_0 -f p^2\right)\tilde{\rho}^0_n\,,\nonumber \\
0 =&& \partial_z\left(\frac{f}{z}\partial_z\tilde{\rho}^\pm_n\right) +\frac{1}{zf}\left[ \left(w_\pm\mp\bar{V}^3_t\right)^2 -fp^2\right]\tilde{\rho}^\pm_n\,,
\end{eqnarray}
where all fields are subject to boundary conditions so that $\rho$-meson fields vanishes at the $z=0$ boundary and the the field strength vanishes at the wall. Before numerically solving these boundary value problems on the constant density, i.e. the fixed value of $Q$, we need to recall the density dependence of the deconfinement phase transition \cite{Lee:2013oya}.
\begin{figure}
\includegraphics{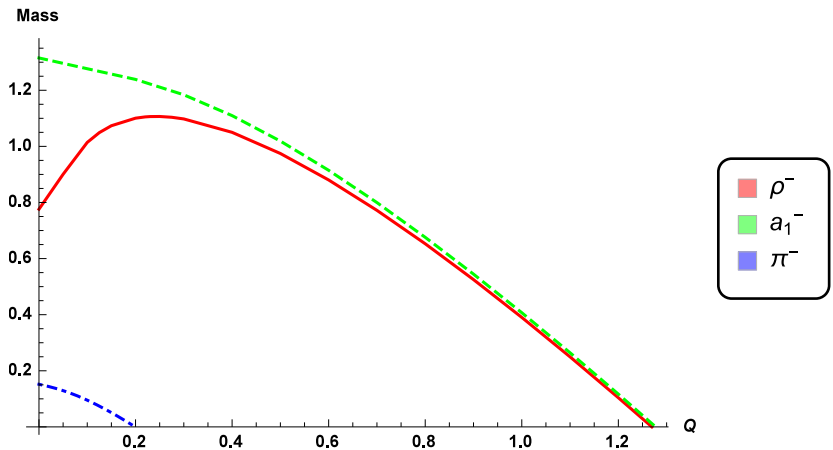}
\caption{Density dependence of mass spectra of negative Mesons\label{DensityDependMass} : $\alpha=1/2$ and mass in GeV units. there are critical density for vanishing meson mass. red line represents for $\rho^-$, green dashed line for $a^-$, and blue dot-dashed line for $\pi^-$.}
\end{figure}

It is shown in the FIG. \ref{DensityDependMass} that each negative meson has its own critical density where the meson becomes massless and a deconfinement phase transition occurs. This critical density is affected by the background density, which is controlled by the parameters $Q$ and $\alpha$. For example, when $\alpha=\frac{1}{2}$, the critical densities of the pion, the vector meson, and the axial-vector meson are $Q_\pi=0.1993$, $Q_\rho=1.269$, and $Q_{a_1}=1.277$, respectively. However, the critical density for the Hawking-Page deconfinement phase transition occurs at $Q_{crit}=0.1679$ and this is smaller than $Q_\pi$. To study mesons in the confining phase, it is needed to keep such density-parameters $Q$ and $\alpha$ as to be fixed in the appropriate range. In this paper, we fix $\alpha=0.5$ and either $Q=0.1$ or $Q=0.01$.

\begin{figure}
\subfigure[\,{\rm Dispersion relation in the dense medium : red real line represents for $\rho^0$, green dashed line for $\rho^+$, blue dot-dashed line for $\rho^-$, and cyon dotted line for mesons at $Q=0$}]{\includegraphics[width=0.45\textwidth]{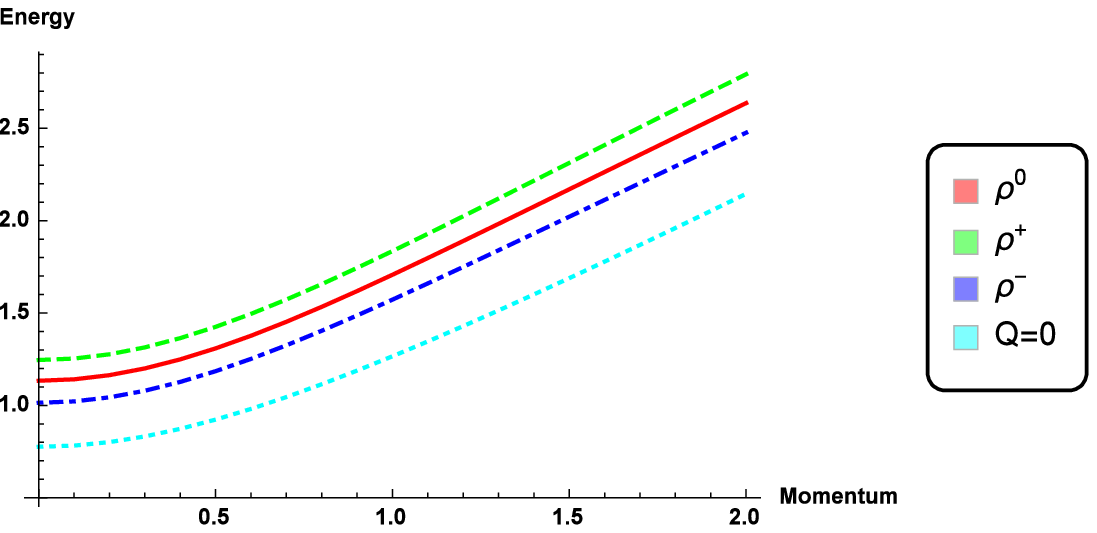}}\hfill
\subfigure[\,{\rm Energy Differences in the dense medium : red line is energy difference of $\rho^+$ and $\rho^0$, and green dashed line one of $\rho^0$ and $\rho^-$}]{\includegraphics[width=0.45\textwidth]{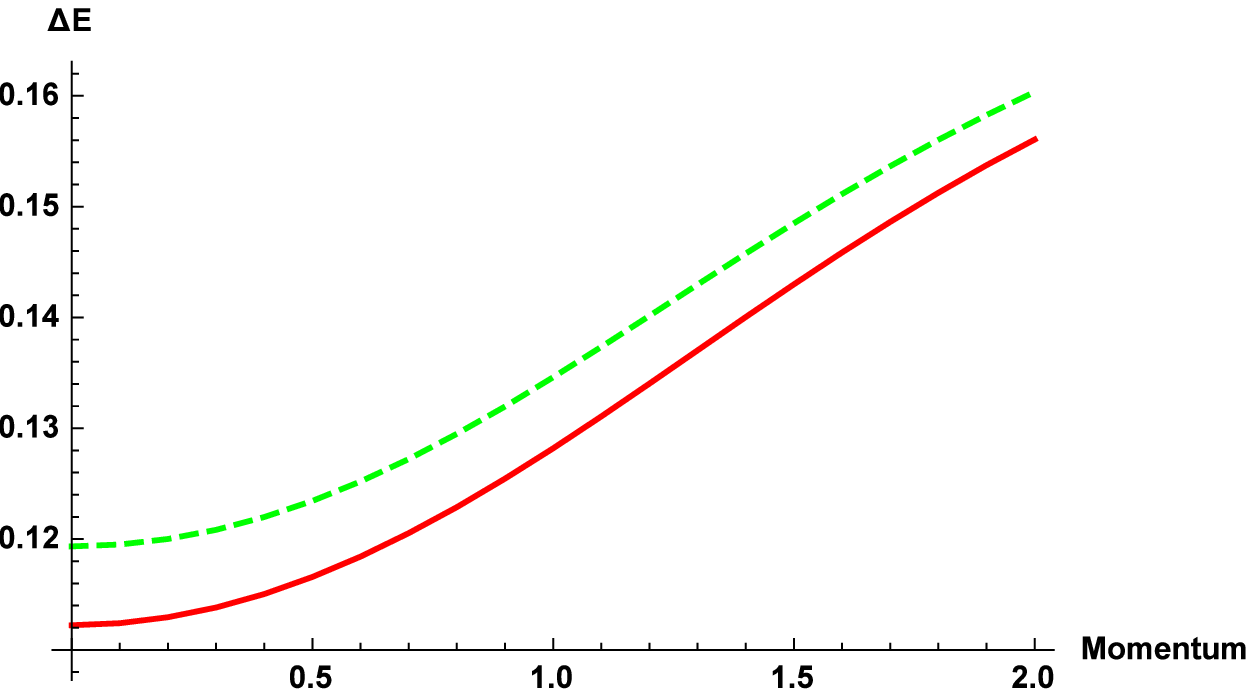}}\\
\caption{Dispersion relations of the $\rho$ mesons at $Q=0.1$ and $\alpha=0.5$\label{DR-rho}}
\end{figure}
We begin to investigate dispersion relations of rho mesons. As shown in the FIG. \ref{DR-rho}, the energies of $\rho$ mesons on the dense medium grow up as their spatial momenta $p$ increase. At a given momentum, the energies of neutral and charged rho mesons split and this results from the existence of the non-zero nuclear density. In general, the energy curve of the $\rho^+$ is placed over the curve of the $\rho^0$ and the curve of the $\rho^-$ is placed at the bottom. Moreover, the splitting among the curves become large in the heavier dense medium.  Without the background densities, energies of all neutral and charged mesons grow up again as the momenta increase, but there is no splitting between the rho mesons. This is the expected result because we know that mesons have different masses according to their charges, at a given non-zero density. Note that masses of the neutral and the positive rho meson on the dense medium are always bigger than those on the vacuum, but the case of the negative rho meson is dependent of the nuclear densities. As in the FIG. \ref{DensityDependMass}, the mass of the $\rho^-$ on the vacuum reads $0.77603\, GeV$, which is the same mass evaluated at the $Q=0.69636$. If we consider a dense medium given by $Q\geq0.69636$, then energy curves of the negative $\rho$'s on the vacuum and the dense medium intersect each other, because the gradients of the curves for the vacuum and the dense medium differ in the region of the higher momentum. In the regime of high spatial momenta, the energy curve of the $\rho^-$ on a dense medium has greater gradient than that on the vacuum. But this does not happen at the confining phase, because the isospin density for the confining phase is always lower than the critical density. It is also noted that energy differences between rho mesons at a given momentum are not same but asymmetric as shown in Fig. \ref{DR-rho} (b).

\begin{figure}
\subfigure[\,{\rm Decay constants in the dense medium : while the red real line represents for $\rho^0$, the green dashed line for $\rho^+$, the blue dot-dashed line for $\rho^-$ at $Q=0.1$, but the cyon dotted line for rho mesons at $Q=0$}]{\includegraphics[width=0.45\textwidth]{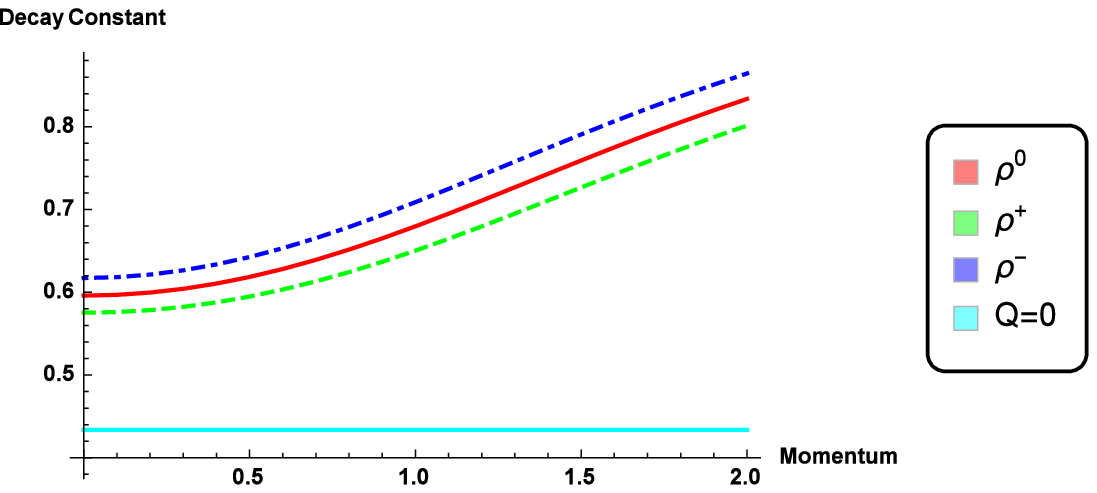}}\hfill
\subfigure[\,{\rm Difference of Decay constants in the dense medium : the red line is decay-constant difference of $\rho^+$ and $\rho^0$, and the green dashed line one of $\rho^0$ and $\rho^-$}]{\includegraphics[width=0.45\textwidth]{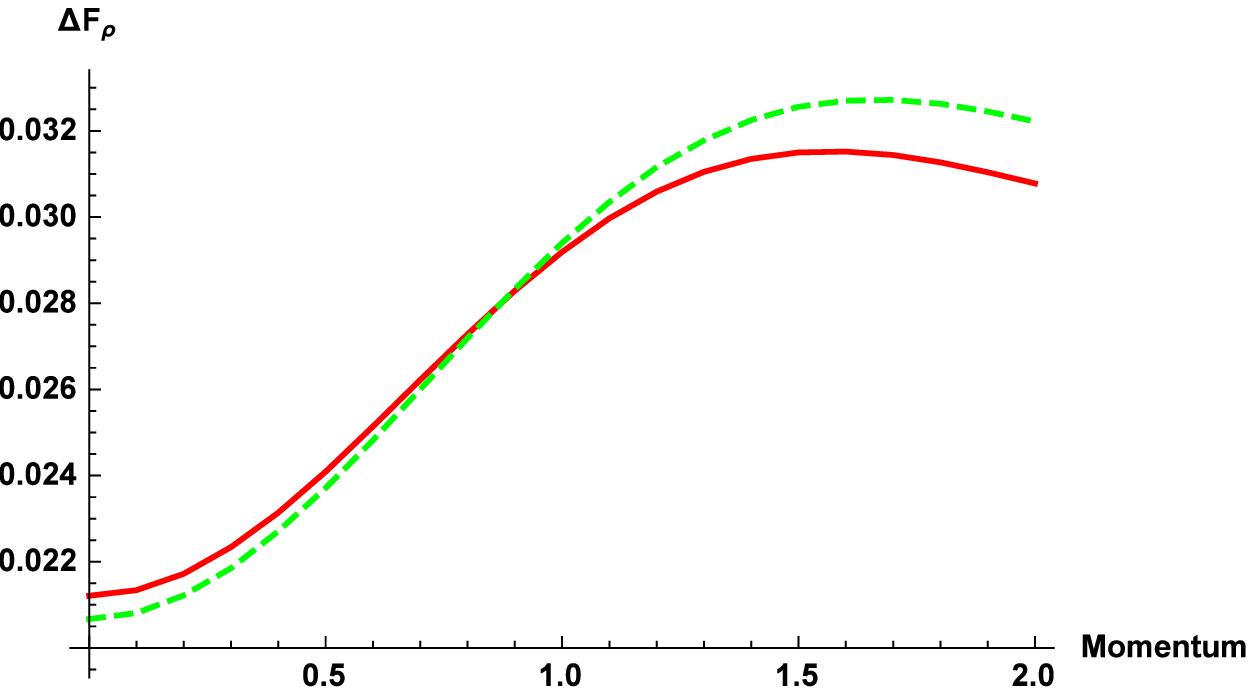}}\\
\caption{Decay constants of the $\rho$ mesons at $Q=0.1$ and $\alpha=0.5$\label{DC-rho}}
\end{figure}
The decay constant of the vector meson can be defined by calculating the current-current correlator in the gague/gravity duality \cite{Erlich:2005qh, Grigoryan:2007vg}. By using the fact that Green's function of bulk gauge field can be expanded in resonances, the vector current-current correlator is derived as a sum over rho mesons \cite{Erlich:2005qh, Erlich:2012HUGS}, which can be explicitly shown in \cite{Grigoryan:2007vg} by incorporating the Kneser-Sommerfeld expansion with the known exact solutions. The suggested form of the decay constant is given by
\begin{equation}
F^2_\rho = \left.\frac{1}{g}\frac{\partial_z \rho}{z}\right|_{z=\epsilon\rightarrow 0}\,,
\end{equation}
where the normalized rho vector field is used. We evaluate the formula numerically and the results are presented in the FIG. \ref{DC-rho}. In the case of the vacuum, decay constants of all the rho mesons have the same value and are independent of their spatial momenta. The numerical value of the decay constant reads $0.43365$ at $Q=0=D$. But, on the dense medium, the non-trivial momentum-dependence of the decay constant is developed. Like in the case of the energy growth, the value of the decay constant becomes larger as the spatial momenta grow up. Decay constants of the neutral and the charged rho mesons have different values and they split more widely as $Q$ increases. Among the curves of the decay constants, the curve for the $\rho^-$ is located at the top and the curves for the $\rho^0$ and the $\rho^+$ are followed successively. At a glance, it seems to be peculiar that the decay constant of the more energetic $\rho^+$ meson is smaller than that of the less energetic $\rho^-$ meson. The naive explanation for this may be given by the dependence of the transition rate on the chemical potentials, correspondingly on the background of excessive isospin chemical potential. Note that the parameter $\alpha=0.5$ represents the isospin matter consisting of $75\%$ proton and $25\%$ neutron. This excessive background charge makes the decay of a meson of the same charge slow and the decay constant of the negative rho-meson is larger than that of the positive rho-meson.



\section{Dispersion Relation and Decay Constant of the $a_1$ Meson}
As like as the case of the vector gauge field, we consider the lowest mode of the axial-vector gauge field and identify the transverse parts of the decomposition (\ref{Decomposition}) as spin-$1$ $a_1$ mesons by field-redefinitions,
\begin{equation}
a_{1m}^0 ~=~ \bar{a}^3_m~~~~~\textrm{and}~~~~~a_{1m}^{\pm} ~=~ \frac{1}{\sqrt{2}}(\bar{a}^1_m\pm i \,\bar{a}^2_m)\,.
\end{equation}
In terms of these, the action governing the $a_1$ axial-vector mesons is given by
\begin{eqnarray}
4g^2\cdot{\cal{L}}_{a} =&& \partial_Ma^0_{1n}\partial^Ma^{0n}_1+4g^2\phi^2a^0_{1m}a_1^{0m}\\
&&+2\left\{\partial_Ma^+_{1n}\partial^Ma^{-n}_1 +\left(\bar{V}_t^3\bar{V}^{3t}+4g^2\phi^2\right)a^+_{1m}a^{-m}_1 +i\bar{V}^3_t\left(a^+_{1m}\partial^ta^{-m}_1-a^-_{1m}\partial^ta^{+m}_1\right)\right\}\nonumber\,.
\end{eqnarray}
The equations of motion are easily written in the four-dimensional momentum space by
\begin{eqnarray}
0 =&& \partial_z\left(\frac{f}{z}\partial_z\tilde{a}^0_{1n}\right) +\frac{1}{zf}\left(w^2_0-fp^2\right)\tilde{a}^0_{1n} -\frac{4g^2\phi^2}{z^3}\tilde{a}^0_{1n}\,,\nonumber\\
0 =&& \partial_z\left(\frac{f}{z}\partial_z\tilde{a}^\pm_{1n}\right) +\frac{1}{zf}\left[\left(w_\pm\mp\bar{V}^3_t\right)^2-fp^2\right]\tilde{a}^\pm_{1n} -\frac{4g^2\phi^2}{z^3}\tilde{a}^\pm_{1n}\,.
\end{eqnarray}
The equations of the charged $a_1$ mesons contain a combination of the energy $w_\pm$ and the density parameters. This combination causes splitting of the physical quantities in the charged $a_1$ mesons such as curves in the dispersion relations and the decay constants. Actually, there exist similar terms in the equations of the motion of charged mesons and this effect can be easily seen without spatial momenta. In comparison with the $\rho$-meson case, the same numerical method is applied to the system of the axial-vector mesons. Again, we impose the Dirichlet boundary condition at the U.V. and the Neumann boundary condition at the I.R. in a gauge-invariant way.

\begin{figure}
\subfigure[\,{\rm Dispersion relation in the dense medium : red real line represents for $a_1^0$, green dashed line for $a_1^+$, blue dot-dashed line for $a_1^-$, and cyon dotted line for mesons at $Q=0$}]{\includegraphics[width=0.8\textwidth]{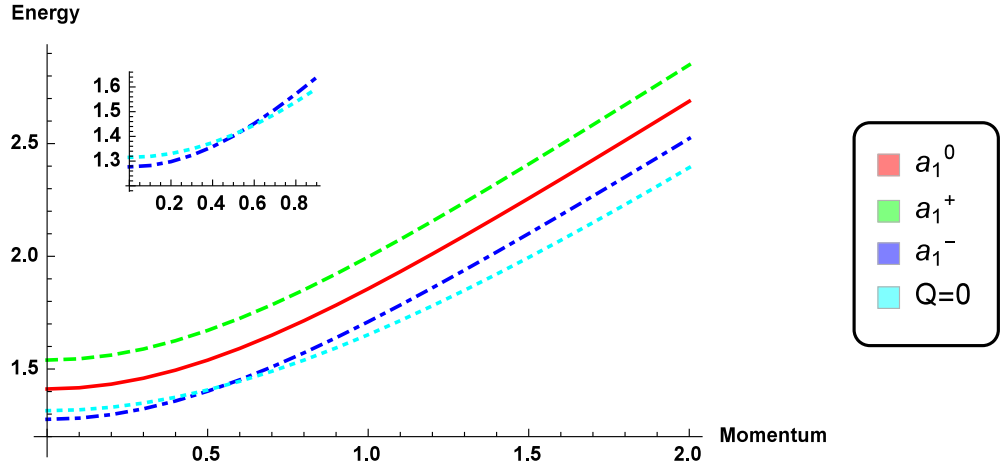}}\\
\subfigure[\,{\rm Energy Differences in the dense medium : red line is energy difference of $a_1^+$ and $a_1^0$, and green dashed line one of $a_1^0$ and $a_1^-$}]{\includegraphics[width=0.7\textwidth]{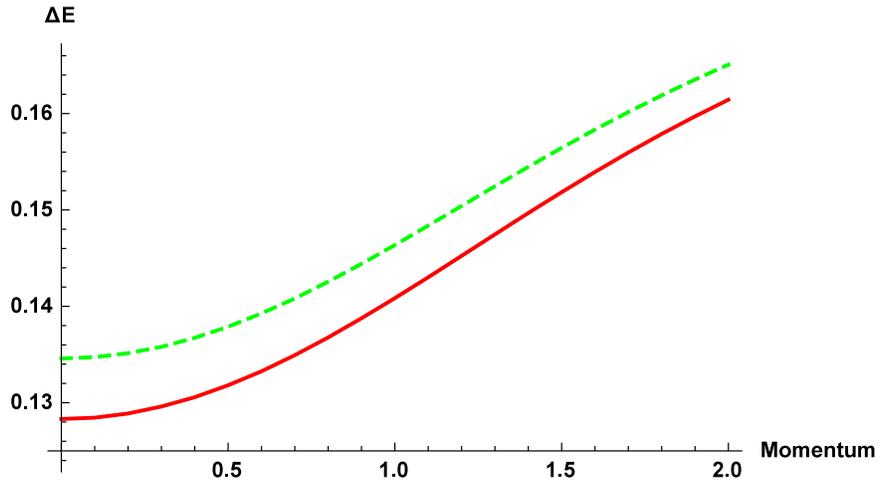}}\\
\caption{Dispersion relation of $a_1$ mesons at $Q=0.1$ and $\alpha=0.5$\label{DP-a}}
\end{figure}
We analyze non-trivial dispersions of the $a_1$ mesons on the dense medium. As in the $\rho$ mesons, the similar results are obtained and presented in the FIG. \ref{DP-a}. The curves for the dispersion relation on the dense medium are located upper than the curve for the vacuum, except for the curve for $a_1^-$ meson. The intersection of the dispersion curves in the negative mesons always exists. But this occurs in a density-independent way for the $a_1^-$ and $\pi^-$ mesons but in density-dependent way for the $\rho^-$ meson. As explained in the section of the $\rho$ meson, the energy of the $a_1^-$ meson on the dense medium becomes large than those on the vacuum at the high momentum region.

\begin{figure}
\subfigure[\,{\rm Dispersion of the positive mesons}]{\includegraphics[width=0.3\textwidth]{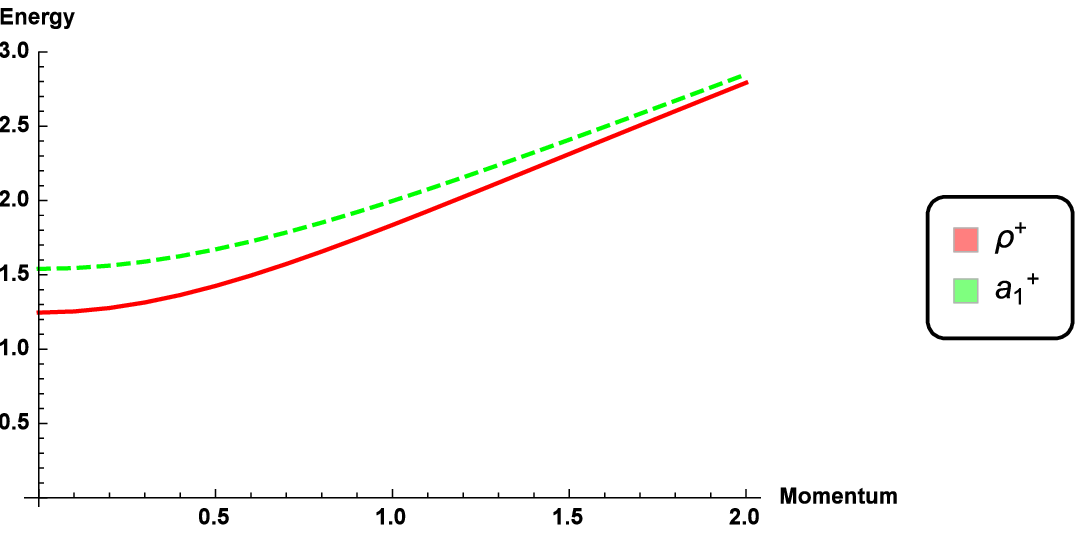}}\hfill
\subfigure[\,{\rm Dispersion of the neutral mesons}]{\includegraphics[width=0.3\textwidth]{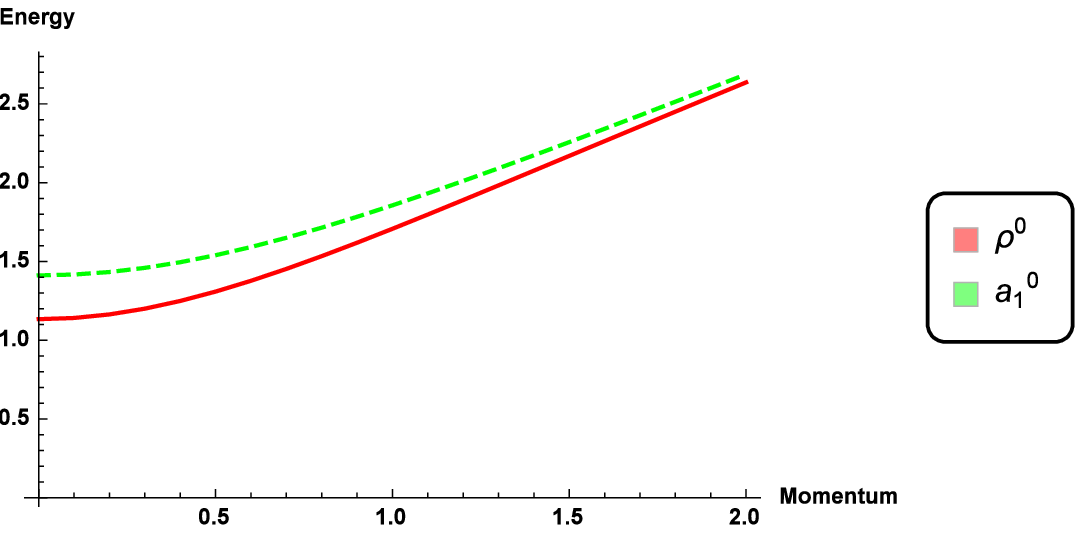}}\hfill
\subfigure[\,{\rm Dispersion of the negative mesons}]{\includegraphics[width=0.3\textwidth]{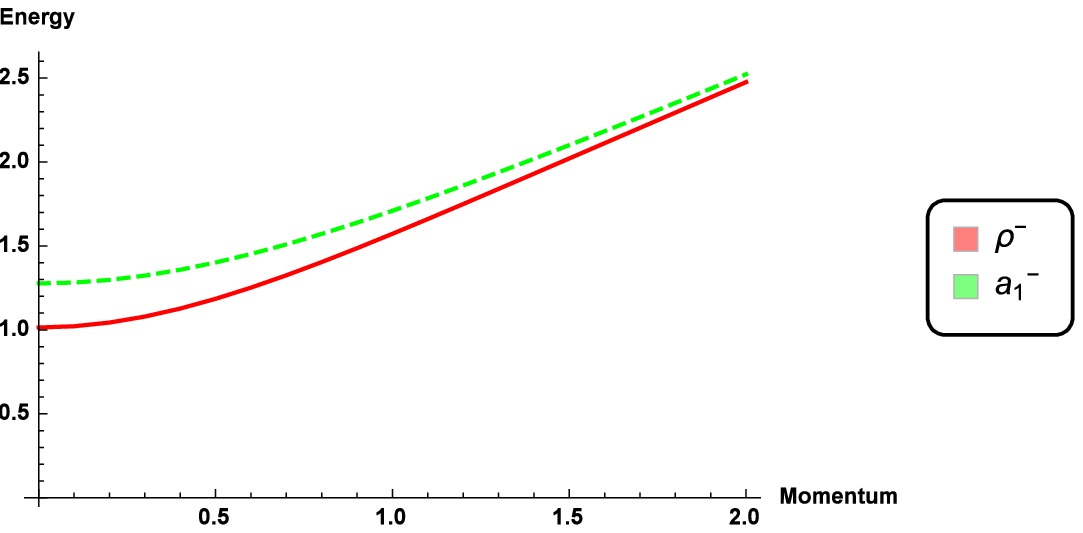}}\hfill
\caption{Dispersion relation of $\rho$ and $a_1$ mesons of the same charge at $Q=0.1$ and $\alpha=0.5$\label{DP}}
\end{figure}
At this point, we compare the curves between the vector and the axial-vector mesons in the FIG. \ref{DP}. The curves for the vector and the axial-vector mesons of the same charge approach each other in the large momenta region. This coincidence implying the same asymptotic behaviour in large momentum region was also observed in \cite{Erlich:2005qh, Grigoryan:2007wn}, just as expected from the equations of motion.

\begin{figure}
\subfigure[\,{\rm Decay constants in the dense medium : the red real line represents for $a_1^0$, the green dashed line for $a_1^+$, the blue dot-dashed line for $a_1^-$ on the dense medium, but the cyon dotted line for mesons at $Q=0$}]{\includegraphics[width=0.45\textwidth]{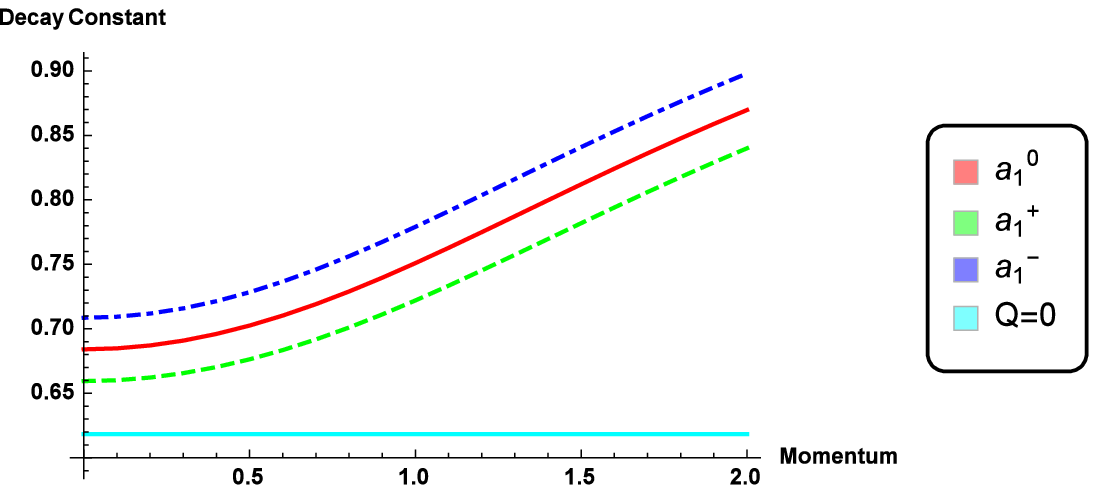}}\hfill
\subfigure[\,{\rm Difference of Decay constants in the dense medium : the red line is decay-constant difference of $a_1^+$ and $a_1^0$, and the green dashed line one of $a_1^0$ and $a_1^-$}]{\includegraphics[width=0.45\textwidth]{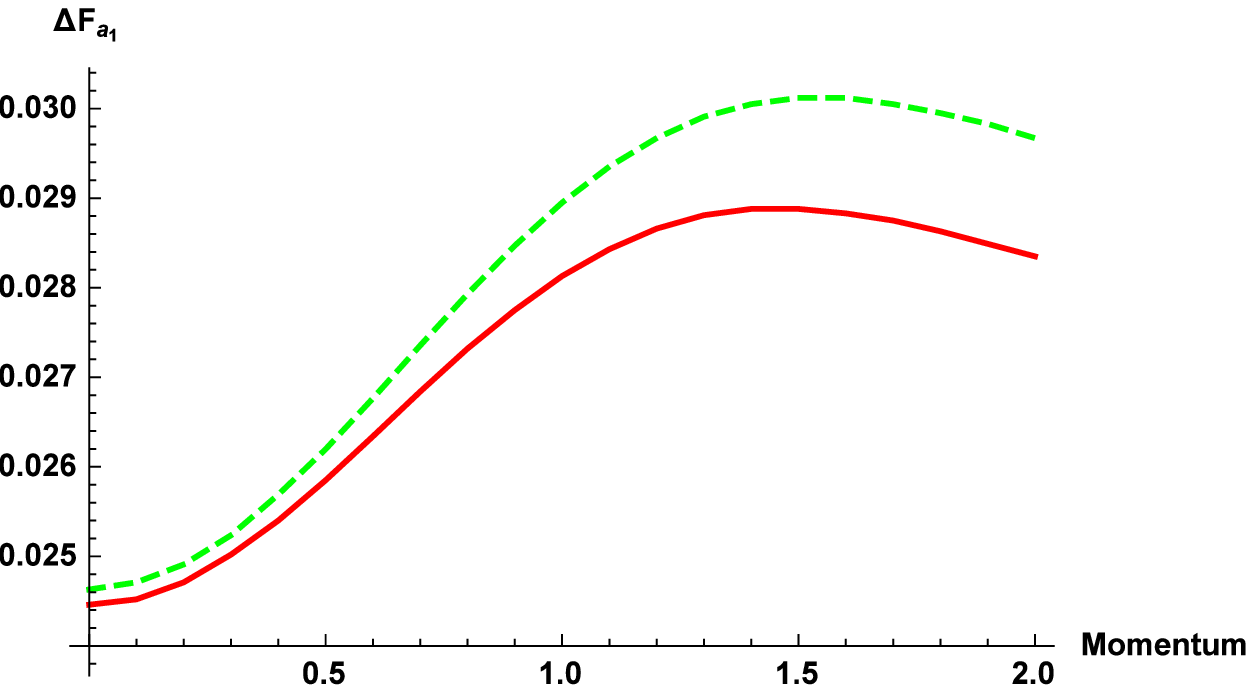}}\\
\caption{Decay constants of the $a_1$ mesons at $Q=0.1$ and $\alpha=0.5$\label{DC-a}}
\end{figure}
Analyzing decay constants of the axial-vector mesons are done likewise in the case of the $\rho$ meson. Numerically, we evaluate the same formula for the decay constant of the vector meson, in this case with the vector field replaced by the normalized $a_1$ fields. The result is shown in the FIG. \ref{DC-a}. Without the isospin chemical potentials, the decay constants of all the $a_1$ mesons have the same constant value $0.61835$, which is bigger than the decay constants of the $\rho$ mesons. However, on the dense medium, the decay constants of the $a_1$ mesons begin to grow up as $p$ increases. The $a_1^+$ meson has the smallest value of the decay constant compared to the others, which is same phenomena as in the case of the rho mesons.

\begin{figure}
\subfigure[\,{\rm Decay constants of the positive mesons}]{\includegraphics[width=0.3\textwidth]{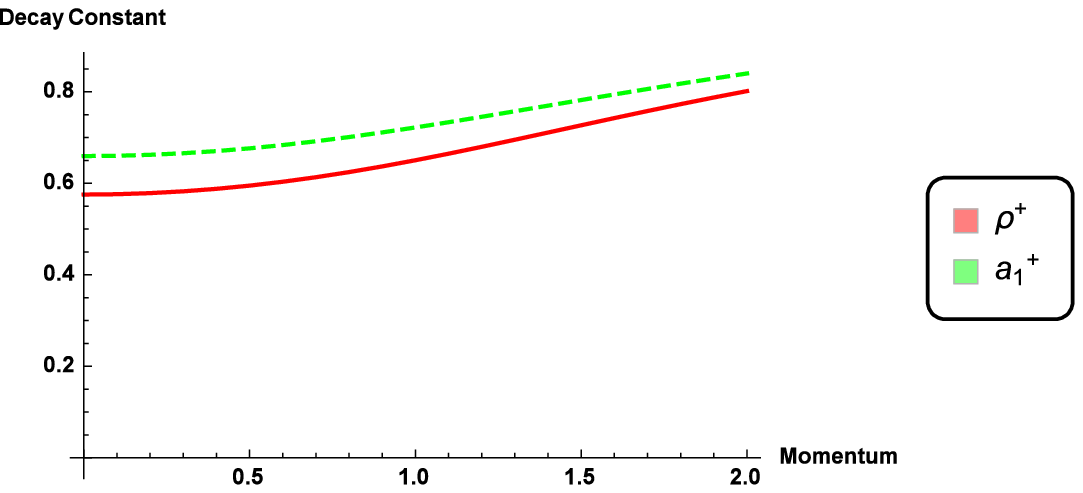}}\hfill
\subfigure[\,{\rm Decay constants of the neutral mesons}]{\includegraphics[width=0.3\textwidth]{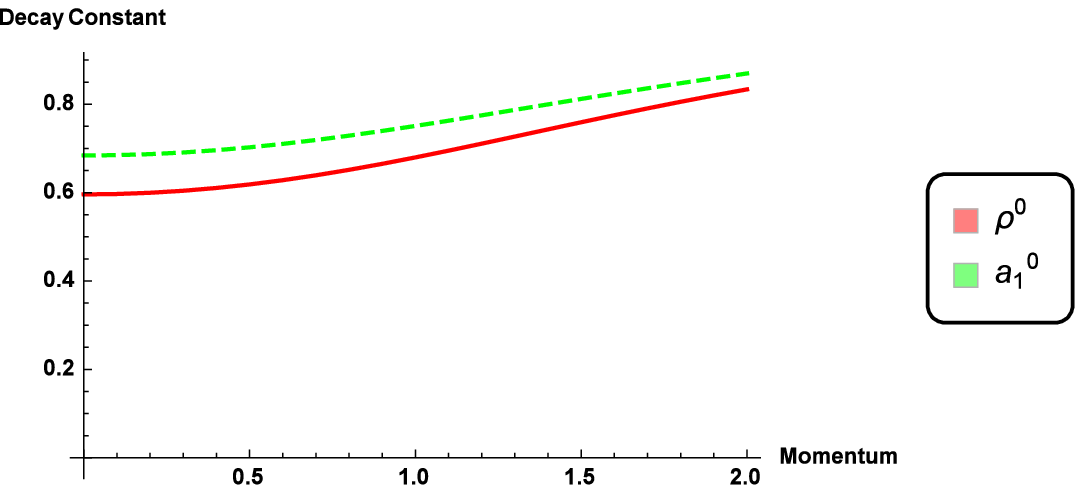}}\hfill
\subfigure[\,{\rm Decay constants of the negative mesons}]{\includegraphics[width=0.3\textwidth]{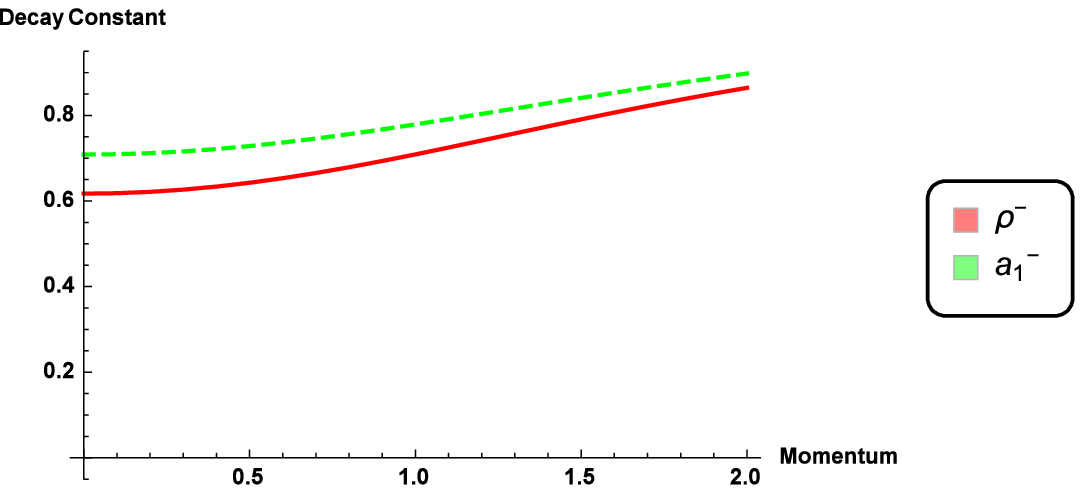}}\hfill
\caption{Decay constants for $\rho$ and $a_1$ mesons of the same charge\label{DC}}
\end{figure}
Like in the case of dispersion relation, we compare the decay constants of the vector and the axial-vector mesons of the same charge in FIG. \ref{DC}. As in the dispersion relation of the energy, the curves of the decay constants for the $\rho$ and the $a_1$ meson of the same charge approach each other in the region of the large spatial momenta. At large $p^2$ region, the asymptotic values of decay constants are of the similar order. Moreover, in each case, the decay constant of the rho meson is smaller than that of the $a_1$ meson of the same charged type, as expected.



\section{Dispersion Relation and Decay Constant of the $\pi$ Meson}

To interpret the lowest modes of the scalar fields as the $\pi$ meson, we make field-redefinitions given by
\begin{eqnarray}
&&\pi^0 = \pi^3\,,~~~~~~~~~~\pi^\pm = \frac{1}{\sqrt{2}} (\pi^1 \pm i \pi^2)\,,\nonumber\\
&&\chi^0 = \chi^3\,,~~~~~~~~~~\chi^\pm = \frac{1}{\sqrt{2}} (\chi^1\pm i\chi^2)\,.
\end{eqnarray}
Then, the Lagrangian for the pseudo-scalar fields is recasted by
\begin{eqnarray}
4g^2\cdot{\cal{L}}_{\chi\pi}
=&&\partial_z\partial_\mu\chi^0\partial^z\partial^\mu\chi^0 +4g^2\phi^2\partial_\mu\chi^0\partial^\mu\chi^0 +2\partial_z\partial_\mu\chi^+\partial^z\partial^\mu\chi^- +8g^2\phi^2\partial_\mu\chi^+\partial^\mu\chi^- \nonumber\\
&& +2\bar{V}^3_t\bar{V}^{3t}\partial_m\chi^+\partial^m\chi^- +4g^2\phi^2\partial_M\pi^0\partial^M\pi^0 +8g^2\phi^2\left\{\partial_M\pi^+\partial^M\pi^- +\bar{V}^3_t\bar{V}^{3t}\pi^+\pi^-\right\}\nonumber\\
&&+8g^2\phi^2\left\{i\bar{V}^3_t\left[ \pi^+\partial^t(\pi^--\chi^-) -\pi^-\partial^t(\pi^+-\chi^+)\right] -\partial_\mu\chi^0\partial^\mu\pi^0\right.\nonumber\\
&&\left. -(\partial_\mu\chi^+\partial^\mu\pi^- +\partial_\mu\chi^-\partial^\mu\pi^+)\right\}\,.
\end{eqnarray}
The equations of motion with respect to the scalar sector consist of the systems of coupled second-order ordinary differential equations,
\begin{eqnarray}
0=&& \partial_z\left(\frac{g^2\phi^2f}{z^3}\partial_z\tilde{\pi}^0\right) +\frac{g^2\phi^2}{z^3f}\left(w^2_0-f p^2\right)\left(\tilde{\pi}^0-\tilde{\chi}^0\right)\,,\nonumber\\
0=&& \partial_z\left(\frac{g^2\phi^2f}{z^3}\partial_z\tilde{\pi}^\pm\right) +\frac{g^2\phi^2}{z^3f}\left[ \left(w_\pm \mp\bar{V}^3_t\right)^2 -f p^2\right]\tilde{\pi}^\pm  -\frac{g^2\phi^2}{z^3f}\left[ w_\pm^2 \mp w_\pm\bar{V}^3_t -f p^2\right]\tilde{\chi}^\pm\,,\nonumber\\
0=&& \partial_z\left[\frac{w_0^2-fp^2}{z}\partial_z\tilde{\chi}^0\right] +\frac{4g^2\phi^2}{z^3f}\left(w^2_0-f p^2\right)\left(\tilde{\pi}^0-\tilde{\chi}^0\right)\,,\nonumber\\
0=&& \partial_z\left[\frac{w_\pm^2-fp^2}{z}\partial_z\tilde{\chi}^\pm\right] +\frac{4g^2\phi^2}{z^3f}\left(w^2_\pm \mp\bar{V}^3_tw_\pm-f p^2\right)\tilde{\pi}^\pm\nonumber\\
&&~~~~~~~~~~~~~~~~~~~~~~~~~~~~~~~~~~~~~~~~~~~~~~~-\frac{4g^2\phi^2}{z^3f}\left[w^2_\pm-fp^2 +\frac{\bar{V}^3_t\bar{V}^3_tz^2}{4g^2\phi^2}p^2\right]\tilde{\chi}^\pm\,.
\end{eqnarray}

\begin{figure}
\subfigure[\,{\rm Dispersion relation in the dense medium : red real line represents for $\pi^0$, green dashed line for $\pi^+$, blue dot-dashed line for $\pi^-$, and cyon dotted line for mesons at $Q=0$}]{\includegraphics[width=0.8\textwidth]{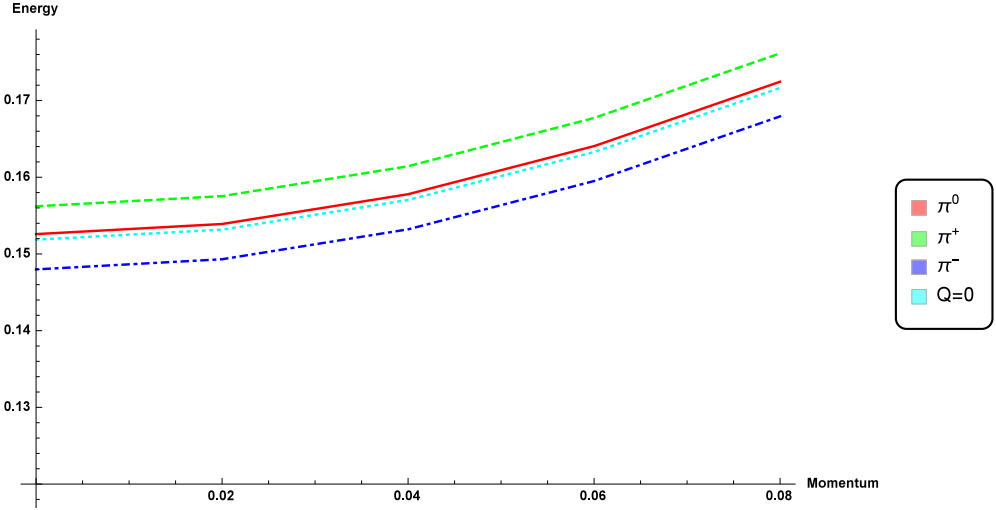}}\\
\subfigure[\,{\rm Energy Differences in the dense medium : red line is energy difference of $\pi^+$ and $\pi^0$, green dashed line one of $\pi^0$ and $\pi^-$, and blue dot-dashed line one of $\pi$ at $Q=0$ and $\pi^-$}]{\includegraphics[width=0.7\textwidth]{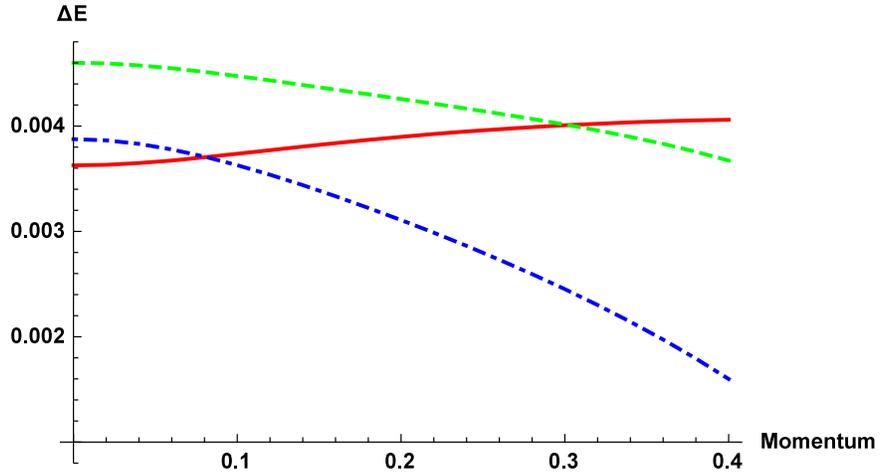}}\\
\caption{Dispersion relation of the pions at $Q=0.01$ and $\alpha=0.5$\label{DR-p}}
\end{figure}

To solve the equations of the scalar mesons, we must specify boundary conditions, which are basically given by the same types used in the case of the vector and axial-vector mesons. The numerical result with $Q=0.01$ and $\alpha=0.5$ is presented in the Fig. \ref{DR-p}. The dispersion curves of the $\pi^+$ and the $\pi^0$ mesons on the dense medium sit over the curve of the $\pi^-$ and the curve of pions at $Q=0$. As expected, the dispersion curves of the $\pi^-$ and the pions at the vacuum intersect each other and the energy of the negative pion on dense medium will be larger than the energy of the pions on the vacuum, seen by the blue dotted-line in the Fig. \ref{DR-p} (c).

We close this section by mentioning difficulty in defining decay constants of pions. In \cite{Erlich:2005qh}, pions are considered as mixture of longitudinal part of the axial vector field $a^i_\mu$ and scalar field $\phi$. Under the chiral limit, the coupled scalar equations are simplified and it is clear what is called pion wave function. Thus, the pion decay constant can be defined by $<0|J^\mu_A|\pi(p)>=i f_\pi p^\mu$ where $J^\mu_a$ is the current for the axial vector field. But, due to the non-trivial metric factors of the dense background, it is hard to find out such a good-looking limit at least to get a numerically calculable form. Moreover, the ground state of dual theory in dense medium is actually not well-defined.



\section{Discussion}

The hard-wall model of the holographic QCD on the dense background was studied with the help of numerical methods. By turning on the spatial momenta $p^2$ in the boundary space, the non-trivial behaviors of the energy dispersions and decay constants of the holographic mesons are developed. As the momenta increase, they increase and, at the same time, split due to the isospin charges. At large $p^2$, the asymptotic values of the energy and decay constants of the $\rho$ and $a_1$ mesons approach each other.

At a given $p^2$, the energy of positive meson is always larger than the energy of the neutral and negative meson of the same type. While the energy of the positive and neural mesons on the dense nuclear medium is always more large than that of the mesons on the vacuum, but the cases for the negative mesons are dependent upon the isospin chemical potential. However, the energy of the meson on the dense nuclear medium is getting bigger than that of the meson on the vacuum in the large $p^2$ regime.

As for the decay constants, the similar patterns were observed : the larger the spatial momenta are, the higher the decay constant of the meson is and the splitting is also existed on the dense nuclear medium. On the other hand, the value of the decay constant of the mesons on the vacuum remains literally 'constant,' irrespectively of the magnitude of the spatial momenta. We also found that, on the dense nuclear matter, the decay constant of the negative meson are always more larger than that of the positive meson of the same type. This was explained by considering the decaying tendency of the charged mesons immersed in the background having isospin chemical potentials.

Since the dense background used for the various fluctuations was not simple, we could not have enough analytic controls and use popular approximations like the chiral limit. While the usual definitions of the QFT in the vacuum are well-known, but, in the dense medium, they are not well-known. So it is hard to study either the decay constant of pion or the meson form factor to investigate the interactions on the dense nuclear medium, which we leave as future works.



\begin{acknowledgments}
Bum-Hoon Lee was supported by the National Research Foundation of Korea (NRF) grant funded by the Korea government (MSIP) (2014R1A2A1A01002306). Chanyong Park was supported by Basic Science Research Program through the National Research Foundation of Korea (NRF) funded by the Ministry of Education (NRF-2013R1A1A2A10057490).
\end{acknowledgments}


\end{document}